\documentclass[a4paper,10pt]{jpconf}
\usepackage{amsmath,amssymb}
\usepackage{iopams}
\usepackage{graphicx}
\usepackage{cite}

\newcommand{\bbR}{\mathbb{R}}

\newcommand\defn[1]{{\normalfont\slshape#1\/}}

\newcommand\bip{\beta}
\newcommand\biA{{}^\beta\mspace{-6mu}A}

\newcommand\biOm{{}^\beta\mspace{-1mu}\Omega}
\newcommand\birOm{{}^\beta\mspace{-1mu}\mathring\Omega}

\newcommand\AD[1][\relax]{#1{a}_{\scriptscriptstyle\mspace{-2mu}\Delta}}

\newcommand\Lap[1][\relax]{\ifx#1\relax  \triangle  \else  \triangle_{\mathrm{#1}}  \fi}

\newcommand\sdual{\mathord{\ast}}

\newcommand\ed{\mathrm{d}}
\newcommand\ii{\mathrm{i}}

\newcommand\sR{\mathcal{R}}

\newcommand\Hil[1][\relax]{\ifx#1\relax\mathcal{H}\else\HilScr#1\relax\fi}
\def\HilScr#1#2\relax{\Hil#1{\mathrm{#2}}}

\newcommand\expect[2][\relax]{\ifx#1\relax\langle\else\mathopen#1\langle\fi #2 \ifx#1\relax\rangle\else\mathclose#1\rangle\fi}

\makeatletter
\def\smallunderarrow@#1#2#3#4{%
	\vtop{\ialign{##\crcr
		\setbox0=\hbox{$\scriptscriptstyle #1$}%
		\hskip0.5\wd0$\m@th\hfil#3#4$\crcr
		\noalign{\nointerlineskip\kern1.3\ex@}%
		$\scriptscriptstyle #1$\ifx#3\displaystyle\scriptsize\else\tiny\fi#2#3\crcr}}}
\newcommand\pback[1][\relax]{\mathpalette{\smallunderarrow@{#1}\leftarrowfill@}}
\newcommand\ppback[1][\relax]{\mathpalette{\smallunderarrow@{#1}\Leftarrowfill@}}
\makeatother

\DeclareMathOperator\Int{int}

\begin{document}

\title{Entropy of generic quantum isolated horizons}
%

\author{C Beetle and J Engle\footnotemark}
\footnotetext{Presenter}

\address{Florida Atlantic University, 777 Glades Road, Boca Raton, FL 33431, USA}

\ead{cbeetle@physics.fau.edu, jonathan.engle@fau.edu}


\begin{abstract}
We review our recent proposal of a method to extend the quantization of \textit{spherically symmetric} isolated horizons, a seminal result of loop quantum gravity, to a phase space containing horizons of \textit{arbitrary} geometry.  Although the details of the quantization remain formally unchanged, the physical interpretation of the results can be quite different.  We highlight several such differences, with particular emphasis on the physical interpretation of black hole entropy in loop quantum gravity.
\end{abstract}

\noindent
The confirmation of the Bekenstein--Hawking entropy formula by Ashtekar, Baez, Corichi and Krasnov \cite{abck1997, dl2004, meissner2004} (ABCK) is one of the triumphs of loop quantum gravity.  The ABCK approach begins by quantizing a classical phase space whose points correspond to spacetimes with inner boundary at a \textit{spherically symmetric} isolated horizon \cite{ak2004, abf1998, ack1999, ashtekar_etal2000}.  The calculation relies on spherical symmetry (just of the intrinsic geometry of the horizon itself) to make the symplectic structure on that phase space, and thus the ensuing quantization, well-defined.

This paper is concerned with an apparent inconsistency in the roles played by spherical symmetry before and after quantization in the ABCK approach.  
Imposing spherical symmetry classically restricts the allowed \textit{bulk} fields such that they induce a round metric on the horizon.  After quantization, however, the bulk spin network states in ABCK are (virtually) generic at the horizon, no different from those allowed on an \textit{arbitrary} 2-surface in loop quantum gravity.  In this sense, the \textit{only} place that spherical symmetry is used at all in deriving the \textit{quantum} theory of ABCK is in making the symplectic structure of the \textit{classical} theory well-defined.

Here, we review a new way \cite{be2010} to justify the ABCK symplectic structure classically, including its crucial Chern--Simons surface term.  Most importantly, our proposed scheme does not rely on spherical symmetry, or indeed on any restriction of the intrinsic horizon geometry, to make the symplectic structure well-defined.  Rather, it allows one simply to quantize the phase space of \textit{all} isolated horizons of a given total area $\AD$.  The intrinsic geometry of the horizon, its shape, \textit{is not fixed a priori}.  This approach renders moot the question of how to impose spherical symmetry at the horizon quantum mechanically, which is the key missing ingredient that would make ABCK a consistent quantization of a spherically symmetric horizon.  Rather, our approach renders the ABCK quantization entirely consistent in another way, namely, by broadening the \textit{classical} picture to eliminate the requirement of symmetry \textit{ab initio}.

\section{ABCK Phase Space and Quantization}

\begin{figure}[t]\centering
\begin{tabular}{ccc}
\raisebox{1.2cm}{\includegraphics[width=5cm]{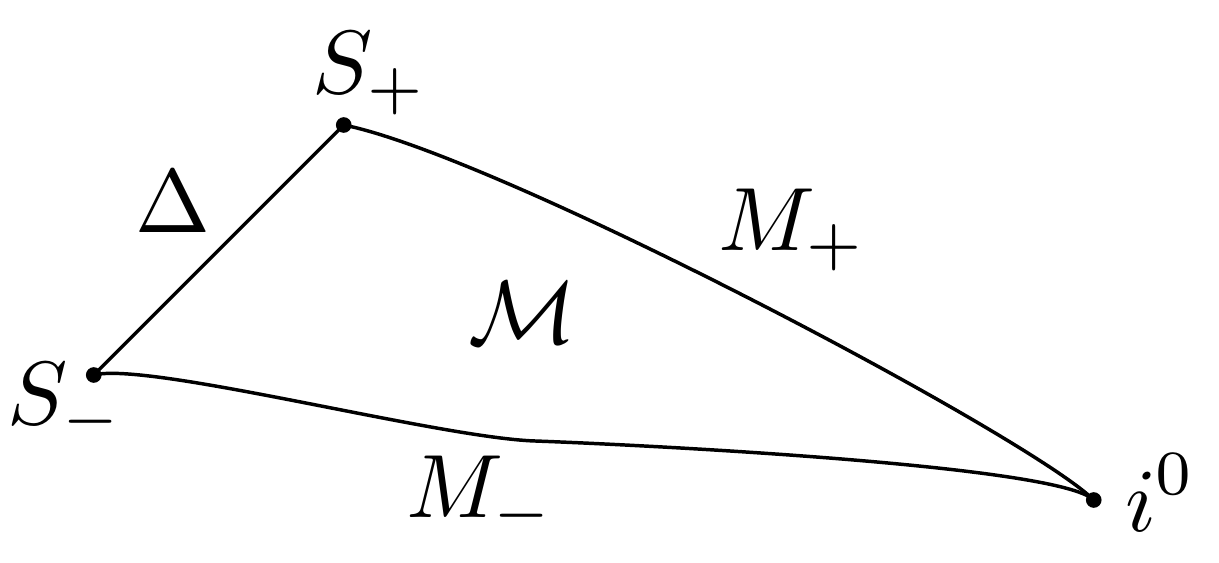}}&\quad
\includegraphics[width=4.5cm]{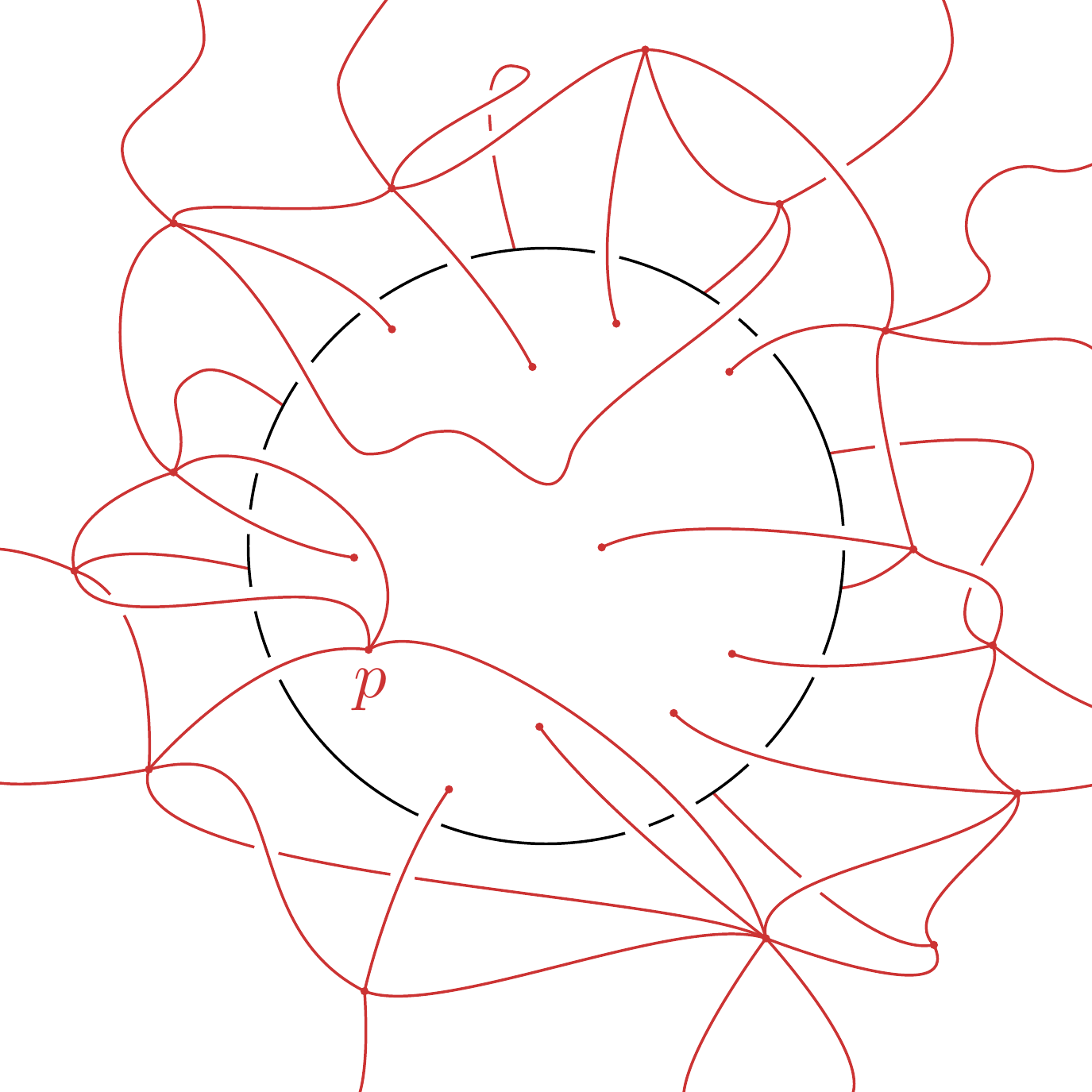}&\quad
\raisebox{1cm}{\includegraphics[width=2.5cm]{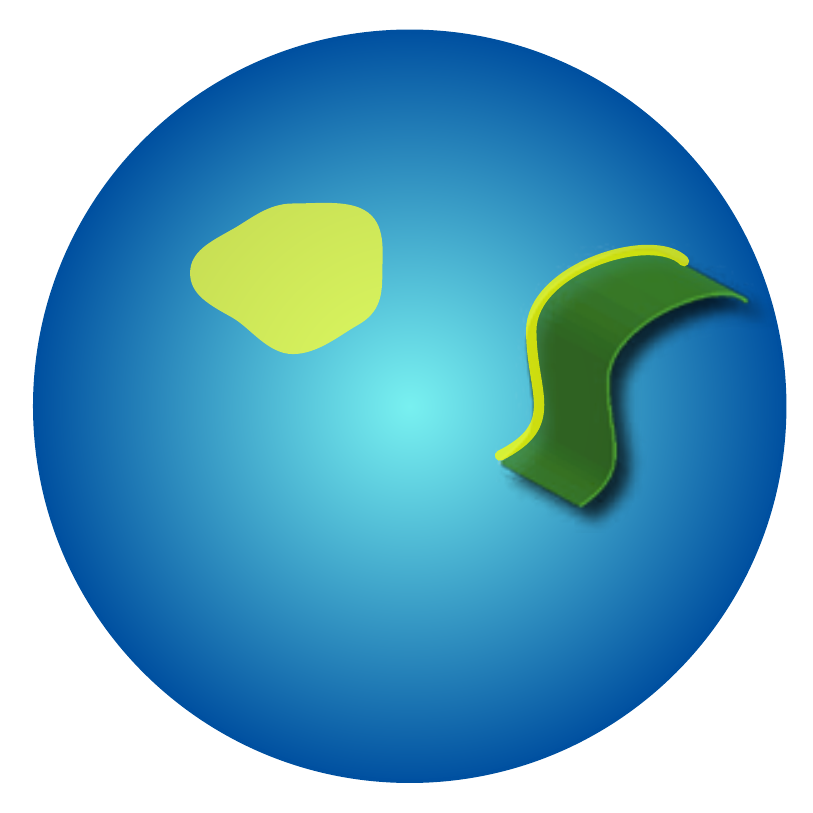}}\\
(a)&(b)&(c)
\end{tabular}
\caption{%
(a) The space-time arena considered both here and in ABCK.
(b) Bulk spin networks fix charges for the quantum Chern--Simons theory via the quantum horizon boundary condition.
(c) Classical horizon shape is probed by transverse, not pullback, fluxes.}
\label{st.sn.tf}
\end{figure}

ABCK use a covariant phase space, each point of which corresponds a space\textit{time} $\mathcal{M}$ of the form shown in Figure \ref{st.sn.tf}(a) in which the classical (Einstein) equations of motion hold.  It is bounded to the future and past by partial Cauchy slices $M_\pm$, which extend to spatial infinity $i^0$, and has an inner boundary at a null surface $\Delta$ diffeomorphic to $S^2 \times \bbR$.  Boundary conditions at $\Delta$ make it is a spherically symmetric \defn{isolated horizon}.  In a precise sense \cite{ak2004}, this means that $\Delta$ models the surface of a quiescent black hole in perfect equilibrium with its immediate surroundings.

Under the above conditions (and one or two additional technical assumptions), the integral
\begin{equation}\label{symInt}
	\Omega(\delta_1, \delta_2)
		:= \biOm_{\mathrm{B}}(\delta_1, \delta_2) + \biOm_{\mathrm{S}}(\delta_1, \delta_2)
		:= \frac{1}{4\pi G \bip} \int_M \delta_{[1} \Sigma^i \wedge \delta_{2]} \biA_i
			+ \frac{1}{2\pi} \frac{\AD}{4\pi G \bip} \oint_{S} \delta_1 V \wedge \delta_2 V
\end{equation}
takes the same value over \textit{any} partial Cauchy slice $M$ through $\mathcal{M}$ with inner boundary $S$.  Here, $\delta_{1, 2}$ represent a pair of tangent vectors to the ABCK phase space, \textit{i.e.}, solutions of the linearized equations of motion on the given background that preserve the total area and spherical symmetry of $S$.  The bulk term $\biOm_{\mathrm{B}}(\delta_1, \delta_2)$  in (\ref{symInt}) is the standard symplectic structure of loop quantum gravity.  The surface term $\biOm_{\mathrm{S}}(\delta_1, \delta_2)$ has a form familiar from a Chern--Simons theory for the $U(1)$ \defn{spin connection} $V_a$ induced on $S$ by the bulk geometry.  The curvature of $V_a$ is
\begin{equation}\label{edV}
	\ed V = - \tfrac{1}{4} \sR\, \epsilon
	\qquad\leadsto\qquad \ed V = -\tfrac{2\pi}{\AD}\, \ppback{\Sigma}_i\, r^i
	\quad \text{if $\sR$ is \textit{constant},}
\end{equation}
where $\sR$ is the the intrinsic scalar curvature of $S$, $\epsilon$ is its 2-from area element, and $r^i$ is a gauge-fixed internal radial direction.  It is the latter equation here, which holds only in spherical symmetry, that makes (\ref{symInt}) independent of $M$, and thus a well-defined symplectic structure.

The sum of bulk and surface terms in the symplectic structure (\ref{symInt}) suggests that the Hilbert space of the quantum theory should be a tensor product $\Hil[^pre] = \Hil[_B] \otimes \Hil[_S]$ of bulk and surface factors.  This is indeed what happens at first in the ABCK quantization.  The bulk Hilbert space $\Hil[_B]$ is the standard one of loop quantum gravity, spanned by spin network states whose underlying graphs may include edges that end at one of a finite set $\mathcal{P}$ of \defn{puncture points} on the horizon.  Such a spin network is shown in Figure~\ref{st.sn.tf}(b).  The surface Hilbert space $\Hil[_S]$ is a direct limit \cite{abck1997} of Hilbert spaces for a quantum Chern--Simons theory, where the limit runs over such sets $\mathcal{P}$ of punctures, ordered by inclusion.  The two Hilbert space factors at this stage are entirely independent of one another and, in particular, the quantum Chern--Simons connection on the horizon is totally unrelated to the geometric degrees of freedom in the bulk.  We denote that connection by $X$, rather than $V$, to emphasize the absence of such a relation.

The initial Hilbert space $\Hil[^pre]$ of the ABCK quantization is reduced to the true, physical Hilbert space $\Hil[^phys]$ of the model in a series of steps.  The first reasserts the physical relationship between the Chern--Simons connection $X$ and the bulk variables by restricting to states $| \psi \rangle$ in $\Hil[^pre]$ that satisfy the \defn{quantum horizon boundary condition}
\begin{equation}\label{qhbc}
	\biggl[ \hat I_{\mathrm{B}} \otimes \hat U[X, C] \biggr] | \psi \rangle
	= \biggl[ \exp \biggl( - \frac{2 \pi \ii}{\AD}\, \hat\Sigma[\Int C, r] \biggr)
		\otimes \hat I_{\mathrm{S}} \biggr] | \psi \rangle.
\end{equation}
Here, $\hat U[X, C]$ is the Chern--Simons holonomy operator around an arbitrary loop $C$ in $S$, while $\hat\Sigma[\Int C, r]$ is the canonical flux of loop quantum gravity, in the gauge-fixed internal direction $r^i$, through the interior of $C$ within $S$.  The quantum horizon boundary condition (\ref{qhbc}) is modeled on the classical relation (\ref{edV}), under the assumption that $\sR$ is \textit{constant}.  The kinematical Hilbert space $\Hil[^kin]$ of states $| \psi \rangle$ obeying (\ref{qhbc}) is then further reduced to the physical Hilbert space $\Hil[^phys]$ of the ABCK model by imposing the diffeomorphism and Hamiltonian constraints.

\section{Area Connection and Quantization of Generic Horizons}

Our proposed application \cite{be2010} of the ABCK quantization to the phase space of \textit{all} isolated horizons with total area $\AD$ hinges on the definition
\begin{equation}\label{accp}
	\mathring V_a := V_a + \tfrac{1}{4} (\sdual \ed\psi)_a
	\qquad\text{with}\qquad
	\Lap\psi := \sdual\ed\sdual\ed\psi := \sR - \expect{\sR}
\end{equation}
of a new $U(1)$ connection on the horizon in the classical theory.  Here, $\Lap$ is the scalar Laplacian and $\sdual$ is the Hodge dual operation, both intrinsic to $S$, while $\expect{\sR} := \oint_S \sR \epsilon / \AD$ denotes the \textit{average} value of the curvature of $S$.  We call the solution $\psi$ of this Poisson equation the \defn{curvature potential} of $S$ and the $U(1)$ connection $\mathring V_a$, the \defn{area connection} because
\begin{equation}\label{edV0}
	\ed \mathring V = - \tfrac{1}{4} \expect{\sR}\, \epsilon = -\tfrac{2\pi}{\AD}\, \ppback{\Sigma}_i\, r^i
	\qquad\leadsto\qquad
	\exp \oint_C \ii \mathring V = \exp \biggl( - \frac{2 \pi \ii}{\AD} \int_{\Int C} \Sigma_i r^i \biggr)
\end{equation}
for \textit{any} geometry.  That is, in close analogy to (\ref{qhbc}), the \textit{classical} holonomy of the area connection about any closed loop $C \subset S$ depends \textit{solely} on the area of $S$ interior to $C$.

Using the area connection $\mathring V_a$ from (\ref{edV0}), we show \cite{be2010} that the symplectic structure integral
\begin{equation}\label{rsymInt}
	\Omega(\delta_1, \delta_2)
		:= \biOm_{\mathrm{B}}(\delta_1, \delta_2) + \birOm_{\mathrm{S}}(\delta_1, \delta_2)
		:= \biOm_{\mathrm{B}}(\delta_1, \delta_2)
			+ \frac{1}{2\pi} \frac{\AD}{4\pi G \bip}
				\oint_{S} \delta_1 \mathring V \wedge \delta_2 \mathring V
\end{equation}
is independent of $M$ throughout the entire phase space of \textit{all} isolated horizons of area $\AD$.  Remarkably, (\ref{rsymInt}) differs from (\ref{symInt}) only in that the surface term involves the area connection rather than the spin connection, and yet is well-defined on a much larger phase space.  Immediately after quantization, \textit{i.e.}, at the level of $\Hil[^pre]$, this classical distinction is irrelevant because the quantum Chern--Simons connection $X$ on $S$ has no physical relation to the bulk variables.  When that relation is restored by the quantum horizon boundary condition (\ref{qhbc}), the meaning given to $X$ is exactly analogous to that of the classical area connection exhibited in (\ref{edV0}).

\section{Quantum Horizon Shape and Entropy}

We now consider the situation from a purely quantum mechanical perspectve.  The mere existence of a classical connection $\mathring V_a$ with holomies (\ref{edV0}) on a generic isolated horizon dispels the notion that just imposing the quantum horizon boundary condition (\ref{qhbc}) makes a quantum horizon spherically symmetric.  Moreover, since (\ref{qhbc}) \textit{does} determine the boundary state uniquely in terms of the bulk, we must look to the bulk to see whether symmetry is actually present.

The geometric content of bulk loop quantum gravity states is probed \cite{al2004} by flux operators
\begin{equation}\label{sfo}
	\frac{1}{8 \pi G \bip}\, \hat\Sigma[T, f] := \frac{\hbar}{2} \sum\nolimits_{x \in T} f^i(x)
		\sum\nolimits_{e \text{ at } x} \kappa(T, e)\, \hat J_i^{(x, e)},
\end{equation}
where the sums are over all analytic curves $e$ extending from each point $x$ of a transversely oriented 2-surface $T$ in $M$, $f^i(x)$ is a Lie-algebra valued smearing function, $\kappa(T, e) = \pm 1$ (or $0$) according to the orientation of $e$ relative to $T$, and $\hat J_i^{(x, e)}$ is a generator of internal $SU(2)$ gauge along $e$. We distinguish between \defn{pullback fluxes}, wherein $T$ is a subset of $S$, and \defn{transverse fluxes}, wherein $T$ intersects $S$ in a curve.  The two cases are illustrated in Figure~\ref{st.sn.tf}(c).

The quantum horizon boundary condition (\ref{qhbc}) involves only pullback fluxes.  Classically, however, information about the intrinsic dyad induced on $S$ inheres in the transverse fluxes.  Because the ABCK quantization yields no restriction on the transverse fluxes, we argue that it is properly viewed as the \textit{consistent} quantization of an isolated horizon of fixed total area $\AD$, but \textit{arbitrary} shape.

In the months since Loops '11, an invitation from Abhay Ashtekar to visit him at Penn State resulted in some very helpful discussions that revealed an important subtlety in our proposed scheme.  Namely, even though we argue that the final Hilbert space contains quantum states corresponding to all possible classical shapes of the horizon, and all of these enter into the statistical ensemble underlying the ABCK entropy, it is \textit{not} these shapes that are counted in determining the entropy. Rather, what one counts are possible states of the ``quantum area element,'' a quantity that classically is \textit{pure gauge.}  Due to  the distributional nature of quantum geometry, it is no longer pure gauge in the quantum theory.

Shortly after this work was finished, a different strategy for allowing arbitrary shapes in the calculation of quantum entropy
was published by Perez and Pranzetti \cite{pp2010}, which, in contrast to the present work, is \textit{$SU(2)$-covariant}.
Let us remark on the relative
strengths and weaknesses
of their work as compared to the present one.
A strength of both
is that quantum black holes \textit{with arbitrary horizon shape} are described,
and an advantage of Perez-Pranzetti over the present approach is that it is $SU(2)$ covariant, something necessary
to obtain the correct coefficient in the next to leading order term in the entropy
\cite{gm2004, gour2002, carlip2000, enp2009, abbdv2009}.
However, a disadvantage of \cite{pp2010} is that one has to perform a separate quantization for each horizon shape (i.e., diffeomorphism equivalence class of horizon geometries) and then combine them,
a procedure which neglects information in the Poisson algebra of observables describing the horizon shape.
As in ABCK, it is also not clear that this fixing of the horizon shape is actually reflected in each  elementary quantization.






\ack

We thank the organizers of the conference for the opportunity to present these results.  We also are indebted to Abhay Ashtekar, who invited us to Penn State to discuss this work.

\section*{References}


\providecommand{\newblock}{}

\newpage

\end{document}